\documentclass[11pt]{revtex4}

\newcommand{\N}{N\raise.7ex\hbox{\underline{$\circ $}}$\;$}

\begin{document}

\title{{\Large
Hawking Radiation in de  Sitter Space: \\
 Calculation of the Reflection Coefficient
  for Quantum Particles
 } }

\author{V. Red'kov}
\author{E. Ovsiyuk}
\author{G.G. Krylov}

 \email{
redkov@dragon.bas-net.by,, e.ovsiyuk@mail.ru, krylov@bsu.by}

\affiliation{
Institute of Physics, NAS of Belarus\\
 Mozyr State Pedagogical University \\
 Belarussian State  University}

%\received{}

\begin{abstract}

Though the problem of Hawking radiation in de Sitter space-time,
in particular  details of penetration of a quantum mechanical
particle through the de Sitter horizon, has been examined
intensively there is still some vagueness in this subject. The
present paper aims to clarify the situation.
 A known
 algorithm for calculation of the reflection coefficient
$R_{\epsilon j}$ on the background of the~de Sitter space-time
model is analyzed. It is shown that the determination of
$R_{\epsilon j}$ requires an additional constrain on quantum
numbers $\epsilon R / \hbar c >> j$, where $R$ is a curvature
radius. When taking into account  this condition, the value of
$R_{\epsilon j}$ turns out to be precisely zero.

 It is shown  that
the basic instructive definition
 for the calculation of  the reflection coefficient in de Sitter model is grounded exclusively on the use of
 zero order approximation  in the expansion of  a particle wave function in a series
 on small parameter $1/R^{2}$,
and it demonstrated that  this recipe cannot be extended on
accounting for contributions of higher order terms. So the result
$R_{\epsilon j}=0$ which has been  obtained  from examining
zero-order term persists and cannot be improved.

 It is claimed
that the calculation of the reflection coefficient $R_{\epsilon
j}$ is not required at all because there is no barrier in the
~effective potential curve  on the background of the~de Sitter
space-time, the later correlate with the fact that the problem in
de Sitter space reduces to a second order differential equation
with only three singular points.  However all known quantum
mechanical problems with potentials containing one barrier reduce
to a second order differential equation with four singular points,
the equation of Heun class.

\end{abstract}

\maketitle

\section{Introduction}

The problem of Hawking radiation [1], and in particular the
radiation in de Sitter space-time [2] and  details of penetration of
quantum mechanical particles through the de Sitter horizon  were
examined in the literature [3--10]. Till now remains some
vagueness in this point, and the present paper aims to clarify the
situation.

In the paper, exact wave solutions for a particle with spin 0 in
 the static coordinates of the de Sitter space-time model
 are examined in detail, and the procedure for calculating  of the reflection coefficient $R_{\epsilon
j}$ is analyzed. First, for scalar particle, two pairs of linearly
independent solutions are specified explicitly:  running and
standing waves.
 A known
 algorithm for calculation of the reflection coefficient
$R_{\epsilon j}$ on the background of the~de Sitter space-time
model is analyzed. It is shown that the determination of
$R_{\epsilon j}$ requires an additional constrain on quantum
numbers $\epsilon R / \hbar c \gg j$, where $R$ is a curvature
radius. When taking into account for this condition, the value of
$R_{\epsilon j}$ turns out to be precisely zero.

 It is claimed
that the calculation of the reflection coefficient $R_{\epsilon
j}$ is not required at all because there is no barrier in the
~effective potential curve  on the background of the~de Sitter
space-time.

The same conclusion holds   for arbitrary particles with higher
spins, it was demonstrated explicitly with the help of the exact
solutions for  electromagnetic and  Dirac fields in \cite{11}.

The structure of  the paper is as follows. In Section 2 we state
the problem;  some more details concerning the approach used
 could be found in \cite{11}.

In Section 3 we demonstrate that the basic instructive definition
 for the calculation of  the reflection coefficient in de Sitter model is grounded exclusively on the use of
 zero order approximation $\Phi^{(0)} (r) $ in the expansion of  a particle wave function in a series
 of the form
\begin{eqnarray}
\Phi (r) = \Phi^{(0)} (r)  + \left ( {1 \over R^{2} }\right  )
\Phi^{(1)}(r) + \left ( {1 \over R^{2}}\right  )^{2} \Phi^{(2)}(r)
+ ...  \label{1.1}
\end{eqnarray}

What is  even more important and we will demonstrate it
explicitly, this recipe cannot be extended on accounting for
contributions of higher order terms. So the result $R_{\epsilon
j}=0$ which will be obtained below from examining zero-order term
$\Phi^{(0)} (r) $ persists and cannot be improved.

\section{ Reflection coefficient}

 Wave equation for a spin 0 particle ($M$ is used  instead of $McR
/\hbar , \; R$ is the curvature radius) reads
\begin{eqnarray}
\left ( \; {1 \over \sqrt{-g} }  \partial _{\alpha}  \sqrt{-g}
 g^{\alpha \beta }\; \partial _{\beta} +  2  +  M^{2}\;
 \right ) \; \Psi (x) = 0\; ,
\label{2.1}
 \end{eqnarray}

\noindent and is considered in static coordinates
\begin{eqnarray}
dS^{2} =   \Phi  dt^{2} -  {dr^{2} \over \Phi } - r^{2} (d\theta
^{2} +  \sin ^{2}\theta  d\phi ^{2})   \;, \qquad
 0 \le  r <  1 \; ,  \qquad  \Phi = 1 -r^{2} \;  .
\label{2.2}
\end{eqnarray}

\noindent  For spherical solutions $ \Psi (x) = e^{-i\epsilon t}
 f(r)  Y_{jm}(\theta ,\phi ) , \; \epsilon  = ER /\hbar c, $
the
 differential equation for $f(r)$ is
 \begin{eqnarray}
 {d^{2} f \over dr^{2}}   + \left  ( { 2 \over r} + {\Phi'  \over \Phi }\right  )  {d f\over dr }
    +  \left   (  {\epsilon ^{2} \over \Phi ^{2} }  -
{M^{2} +2 \over   \Phi }  -    {j(j+1) \over   \Phi r^{2}  } \;
\right  )
 f   = 0     \; .
\label{2.3}
\end{eqnarray}
All solutions are constructed in terms of  hypergeometric
functions (let $r^{2} = z$):

 regular at $r=0$ standing waves are given as
\begin{eqnarray}
f(z) = z^{j/2} \; (1 - z)^{-i\epsilon /2}\; F( a , b , c ; z ) \;,
\qquad
 \kappa = j/2 \; ,\; \;  \sigma = -i\epsilon /2\;, \;\; c =
j + 3/2 \;,
 \nonumber
\\
a =  { 3/2 + j + i \sqrt{ M^{2}-1/4}  - i\epsilon  \over 2} \; ,
\qquad
  b =  { 3/2 + j - i \sqrt{ M^{2}-1/4}  - i\epsilon  \over 2} \; ;
\label{2.5b}
\end{eqnarray}

 singular at $r=0$ standing waves are
\begin{eqnarray}
 g(z)  = z^{-(j+1)/2}\; (1 - z)^{-i\epsilon /2} \; F(
\alpha ,\;\beta ,\; \gamma ; \; z ) \; , \qquad
 \kappa  =
-(j + 1)/2 \; , \;  \sigma  = -i\epsilon /2 \;  , \nonumber
\\
\; c = -j + 1/2 \;, \; \alpha  =  { 1/2 - j + i \sqrt{ M^{2}-1/4}
- i\epsilon  \over 2} \; , \qquad
 \beta = { 1/2 - j - i \sqrt{ M^{2}-1/4}  - i\epsilon
\over 2} \; .
 \label{2.6b}
 \end{eqnarray}

 With the use of the Kummer's  relations, one can expand the standing waves into  linear combinations of
the running waves
\begin{eqnarray}
f(z) =  { \Gamma (c) \Gamma (c - a - b)  \over \Gamma (c - a)
\Gamma (c - b) } \; U^{out}_{run}(z) \; + \; {\Gamma (c) \Gamma (a
+ b - c ) \over \Gamma (a) \Gamma (b) }\;
 U^{in}_{run}(z) \; ,
\label{3.2a}
\end{eqnarray}
\begin{eqnarray}
U^{out}_{run}(z)  = z^{j/2} \; (1 - z)^{-i\epsilon /2} \; F ( a ,
b , a + b - c  + 1 ; 1  -  z ) \;  , \;\; \nonumber
\\
U^{in}_{run}(z)  = z^{j/2}\; (1- z)^{+i\epsilon /2}\;
 F (c- a, c - b, c - a - b + 1; 1 - z) \; ,
\label{3.2b}
\end{eqnarray}
\begin{eqnarray}
a^{*} = ( c - a ) \; ,\;  b^{*} = ( c - b ) \;  ,
\qquad
 ( a + b - c
)^{*}  =  - ( a + b - c ) \; ,
\nonumber
 \end{eqnarray}
\begin{eqnarray}
 [\; U^{out}_{run}(z) \;
]^{*}   =  U^{in}_{run}(z)\; ,
\end{eqnarray}
\begin{eqnarray}
f(z)  =
 2\; \mbox{Re}\;  {\Gamma (c) \Gamma (c - a - b) \over
\Gamma (c - a) \Gamma (c - b)} \; U_{out}(z) =
2\; \mbox{ Re}\;
\;{ \Gamma (c) \Gamma (a  + b  - c ) \over \Gamma (a) \Gamma (b)}
\; U_{in}(z) \; \;  . \label{3.3c}
\end{eqnarray}

\noindent
Similarly for $g(z)$
\begin{eqnarray}
g(z) =   {\Gamma (\gamma ) \Gamma (\gamma  - \alpha  - \beta )
\over \Gamma (\gamma  - \alpha ) \Gamma (\gamma  - \beta )}\;
U^{out}_{run}(z) \; + \; {\Gamma (\gamma ) \Gamma (\alpha  + \beta
- \gamma ) \over \Gamma (\alpha ) \Gamma (\beta ) }\;
U^{in}_{run}(z) \; , \label{3.5a}
\end{eqnarray}
\begin{eqnarray}
U^{out}_{run}(z) = z^{j/2} \; (1- z)^{-i\epsilon /2} F ( \alpha +
1 - \gamma , \beta  + 1 - \gamma , \alpha  + \beta  + 1 - \gamma ;
1  - z )\;  , \nonumber
\end{eqnarray}
\begin{eqnarray}
U^{in}_{run}(z) = z^{j/2}\; (1- z)^{+i\epsilon /2} F ( 1 - \alpha,
1 - \beta ,\gamma + 1 - \alpha - \beta ;1 - z)\; , \label{14.3.5c}
\end{eqnarray}
\begin{eqnarray}
g(z)  = 2 \; \mbox{Re}\; {\Gamma (\gamma ) \Gamma (\gamma -
 \alpha - \beta) \over
 \Gamma (\gamma - \alpha ) \Gamma (\gamma - \beta )}  \;
 U^{out}_{run}(z)=
  2 \; \mbox{Re} \;  {\Gamma (\gamma ) \Gamma (\alpha + \beta  - \gamma  ) \over \Gamma
(\alpha ) \Gamma (\beta ) } \; U^{in}_{run}(z)\;    . \label{3.5d}
\end{eqnarray}

Asymptotic   behavior of  the running waves is given by the relations
\begin{eqnarray}
U^{out.}_{run} ( r \sim  0 ) \sim  { 1 \over r^{j+1}} \; , \;\;
U^{out}_{run}( r \sim  1 ) \sim  (1 - r^{2})^{-i\epsilon /2}\; ,
\nonumber
\\U^{in}_{run}( r \sim 0 ) \sim  {1 \over r^{j+1} }\; ,  \qquad
U^{in}_{run}( r\sim 1) \sim (1 - r^{2})^{+i\epsilon /2}\; ,
 \label{4.2a}
\end{eqnarray}

\noindent or in  new radial variable $r^{*} \in  [ 0,\; \infty  ) $:
\begin{eqnarray}
r^{*}  = {R  \over 2} \; \ln  {1 + r \over 1-r } \;\; , \;\; r
= {\exp (2r^{*}/R ) - 1 \over  \exp (2r^{*}/R ) + 1 } \;\; , \qquad
 \nonumber
\\
U^{out}_{run}( r^{*} \sim  \infty  ) \sim \left (
 2^{-iER /\hbar c} \right ) \;  \exp (+i E r^{*}/\hbar c )\;,\qquad
   \nonumber \\
U^{in}_{run}( r^{*} \sim  \infty  ) \sim  \left ( 2^{+iER
/\hbar c} \right ) \; \exp ( -iE r^{*} /\hbar c ) \;,
\qquad \epsilon =
ER /\hbar c )\; .
\label{4.3b}
\end{eqnarray}

\noindent For the standing waves we have
\begin{eqnarray}
f(r \sim  0) \sim  r^{j}\; ,  \qquad g(r \sim  0) \sim { 1 \over
r^{j+1}} \; , \qquad \qquad  \nonumber
\\
f(r \sim 1 ) \sim   2\; \mbox{Re} \; \left [ \; { \Gamma (c)
\Gamma (a  + b - c ) \over \Gamma (a) \Gamma (b) }\; 2^{+iE R
/\hbar c} \; \mbox{exp}(-iE r^{*} /\hbar c) \;    \right ] ,
\nonumber
\\
g(r \sim 1) \sim     2 \; \mbox{Re}  \;\left [\; { \Gamma (\gamma
) \Gamma (\alpha  + \beta  - \gamma ) \over \Gamma (\alpha )
\Gamma (\beta )  } \; 2^{+iE R /\hbar c} \; \mbox{exp}(-iE
r^{*} /\hbar c) \; \right ] . \label{4.4c}
\end{eqnarray}

On can perform the transition to the limit of the flat space-time in
accordance with the following rules:
\begin{eqnarray}
a = {p + 1 - i \epsilon  R  + i \sqrt{R ^{2} M^{2} - 1/4} \over 2}
\; ,\qquad
 b = {p + 1 - i \epsilon R -i \sqrt{R
^{2}M^{2} -1/4}\over 2}\; , \; p = j + 1/2 \;\; ; \label{4.5}
\end{eqnarray}
\begin{eqnarray}
\lim_{ R \rightarrow  \infty } ( R ^{2} z ) = R^{2}, \;\; F ( a ,
b , c ; z)=
 1  + { a b \over c } {z \over 1!}\;+\; {
a(a+1) b(b+1) \over c(c+1) } \; {z^{2} \over 2!}\; + \; \ldots
\; , \label{4.6a}
\end{eqnarray}

\noindent and further ($R \rightarrow \infty $)
\begin{eqnarray}
{a +n \over R} = {1 \over 2} \left ( {p+1 \over R} - i
\epsilon + i \sqrt{ M^{2} - {1 \over 4 R^{2}}} \right ) + {n
\over R } \approx {-i\epsilon -i M \over 2} \; ,\nonumber
\\
{b +n \over R } = {1 \over 2} \left ( {p+1 \over R} - i
\epsilon - i \sqrt{M^{2} - {1 \over 4 R^{2}}} \right ) + {n
\over R} \approx {-i\epsilon + i M \over 2} \; .
\end{eqnarray}

 \noindent Let $ \epsilon ^{2} - M^{2} \equiv
k^{2} $, then we arrive at
\begin{eqnarray}
\lim _{ R \rightarrow \infty } F (a , b , c ; z ) = \Gamma (1 +
p)\; \sum_{0}^{\infty} {(- k^{2} R^{2}/4)^{n} \over n! \Gamma (1 +
n  + p) }\; , \nonumber \label{4.6b}
\\
\lim_{ R \rightarrow \infty} F(b - c + 1, a - c + 1, - c + 2 ;
z)
 = \Gamma (1 - p) \; \sum_{0}^{\infty} {(-k^{2} R^{2}/4)^{n}
\over n!  \Gamma (1 + n - p)}\; . \label{4.6c}
\end{eqnarray}

 \noindent Allowing for the know expansion for Bessel functions
 \begin{eqnarray}
J_{p}(x) = ({x \over 2})^{p} \; \sum _{0}^{\infty} {(ix/2)^{2n}
\over n! \Gamma ( 1 + n + p)} \; ,
\nonumber
\end{eqnarray}
we  get (wave amplitude $A$ will be determined below)
\begin{eqnarray}
 \qquad  \lim _{ R  \rightarrow
\infty } A\; U^{out}_{run}(z) = \lim_{ R \rightarrow  \infty }
A \;  {1 \over \sqrt{r}}, \nonumber
\end{eqnarray}
\begin{eqnarray}
\times  \left [ {\Gamma  (-i\epsilon R + 1) \Gamma(-p) \;
\Gamma (1+p)  (2/k)^{p}    R^{-p+1/2} \over \Gamma [{1 \over 2}
(+i \sqrt{R ^{2} M^{2} -1/4 } - i\epsilon R + p + 1) ]
\Gamma [{1\over 2} (-i \sqrt{R^{2} M^{2}  - 1/4} - i \epsilon
R + p +1)]} \;\; J_{p}(kr) \right. \nonumber
\\
\left.   + {\Gamma (-i\epsilon R +1) \Gamma(+p)  \Gamma (1-p)
 (2/k)^{-p}    R ^{+p+1/2} \over \Gamma [{1
\over 2} (+i \sqrt{R^{2} M^{2} -1/4} - i\epsilon R -p +1)]
\Gamma [{1 \over 2} (-i \sqrt{R^{2} M^{2} -1/4} - i\epsilon
R -p + 1)]} \;\; J_{-p}(kr)  \right] .
\nonumber
\\
\label{4.8}
\end{eqnarray}

Performing  limiting procedure  (see the detail in [11]) we
derive the relation
\begin{eqnarray}\label{4.12a}
\lim _{ R \rightarrow  \infty } A \; U_{out}(z)   \rightarrow
{ 1 \over i^{j + 1} }  \; \sqrt{{2 \over kr}}\; H^{(1)}_{j + 1/2}
(kr) \;,
\end{eqnarray}
where
\begin{eqnarray}
 H^{(1)}_{j+1/2}(x)  =  { ip \over \sin (\pi p)}
\; \left  [ \; e^{ip\pi } \; J_{p}(x) \; - \; \nonumber
J_{-p}(x)\;\right]
\end{eqnarray}
%$H^{(1)}_{j+1/2}(kr)$
stands for Hankel spherical functions.

In connection with the limiting procedure, let us pose a question:
when the relation (\ref{4.12a}) gives us with a good
approximation provided the curvature radius  $R$ is finite.

This point is important, because when calculating  the reflection
coefficient in the de Sitter space just this approximation
(\ref{4.12a}) was used [3--6].

To clarify this point, let us compare the radial equation in  Minkowski  model
\begin{eqnarray}
\left [ {d^{2}\over dr^{2} } \;+\; {2 \over r } {d \over dr} \;+\;
\epsilon ^{2} \;-\; M^{2} \; - \; {j(j + 1) \over r^{2}} \right ]
 f_{\epsilon j} ^{0} = 0 \; ,
\label{4.15}
\end{eqnarray}
and the appropriate equation in de Sitter model
\begin{eqnarray}\label{rds}
\left [{d^{2}\over dr^{2} } \;+\; {2(1 - 2r^{2}/R^{2} ) \over
r(1 - r^{2}/R^{2})}\; {d \over dr}\; +\; {\epsilon ^{2} \over
(1 - r^{2}/R^{2})^{2} }\; -\; {M^{2}  + 2 \over 1 -
r^{2}/R^{2} } \; -\; {j(j + 1) \over r^{2}(1 - r^{2}/R^{2})} \right ]
f_{\epsilon j}  = 0 \; .
\end{eqnarray}

\noindent
At the region far from the horizon
$r \ll  R $,
the last equation reduces to
\begin{eqnarray}
\qquad \left [ {d^{2}\over d R^{2} } \;+\; {2 \over R }
{d \over d R} \;+\; \epsilon ^{2} \;-\; M^{2}  - {j(j + 1) +2 \over
R ^{2}} - \; {j(j + 1) \over R^{2}} \right ]
 \bar{f}_{\epsilon j} = 0\; .
\label{4.14}
\end{eqnarray}

\noindent So, we immediately conclude that eq.~(\ref{4.15}) coincides
with (\ref{4.14})  only for solutions with quantum numbers  obeying
the following restriction
\begin{eqnarray}
\epsilon ^{2} \; - \; M ^{2}  \gg { j ^{2} \over R^{2}
}\;  . \label{4.18}
\end{eqnarray}

In usual units, this inequality reads
\begin{eqnarray}
E =\mu \; mc^{2} , \qquad \lambda = {\hbar \over mc} \; , \qquad
{\lambda ^{2} \over  R^{2}} \sim 10^{-80}\;, \qquad  \mu^{2} -1
\gg {\lambda ^{2} \over  R^{2}} \;  j^{2} \; . \; \label{B}
\end{eqnarray}

\noindent Instead, in massless case we  have
\begin{eqnarray}
{R^{2} \omega^{2} \over c^{2}} \gg   j^{2}   \qquad\mbox{or} \qquad
{R^{2 } 4\pi^{2} \over \lambda^{2}}  \gg  j^{2}  \; .
\label{B'}
\end{eqnarray}

 So we can state that the above relation  (\ref{4.12a})
is  a good approximation at a finite $R$ only for quantum numbers
obeying (\ref{B})--(\ref{B'}).

One additional point should be emphasized,  the radial equation
in de Sitter space can be transformed to the form of the
Schr\"{o}dinger like equation with an effective barrierless potential.
Indeed, in the variable $r^{*}$   eq.~(\ref{rds}) reduces to
\begin{eqnarray}
\left [ {d^{2} \over dr^{*2}}  +   \epsilon ^{2} - U(r^{*}) \right
]  G(r^{*}) = 0 \; , \qquad \qquad \nonumber
\\
U(r^{*}) =    {1 - r^{2} \over R ^{2}} \left [ 4 (1 - r) \;+\;
{r \over 1 + r} \; + \; m^{2} R^{2} \;+\; {j (j + 1) \over
r^{2}}    \right ] \; . \label{6.1b}
\end{eqnarray}

\noindent It is easily verified that this potential corresponds to
attractive force in all space
\begin{eqnarray}
F_{r^{*}} \equiv    - {d U \over dr^{*} } = {1 - r^{2} \over R
^{2}} \;+\; \left [ 2 r \left ( {j (j + 1) \over r^{2} }\;+\;
m^{2} R ^{2} + \; {r\over 1 + r} \; \right.  \right. \nonumber
\\
\left.  \left.  +\; 4 (1 - r) \right ) \;+\;   (1 - r^{2}) \left (
{2 j (j + 1) \over r^{3} } \;+\; 4 - {1 \over (1 + r)^{2}} \right
) \right ] > 0 \; . \nonumber
\end{eqnarray}

\noindent At the horizon, $r^{*} \rightarrow \infty$,  the  potential
function  $U(r^{*})$ tends to zero, so  $ G(r^{*}) \sim  \exp  (
\pm  i \epsilon r^{*})$.

The form of the  effective Shcr\"{o}dinger equation modeling a
particle in the de Sitter space indicates that the problem of calculation
of the reflection coefficient  in the system should not be even
stated. However, in a number of publications such a problem has been
treated and solved.  So we should reconsider these calculations and results
obtained. Significant steps of our approach are given below:

The existing in  literature calculations of non-zero reflection
coefficients  $R_{\epsilon j}$ were based on the usage of the approximate
formula for $U^{out}_{run}(R)  $ for the region far from horizon (\ref{4.8}).
However, as noted above, the formula used is a good approximation only for
solutions specified by (\ref{4.18}), that is when $
  j \ll \epsilon R  $.

It can be shown  that the formula existing in the literature gives a
trivial result  when taking into account
this restriction. The scheme  of calculation   (more detail see in [11]) is described below.

The first step consists in the  use of the asymptotic formula for the
Bessel functions: when  $x \gg \nu ^{2}$ we have
\begin{eqnarray}
  J (x) \sim  {\Gamma (2 \nu + 1)\; 2^{-2\nu
-1/2} \over
 \Gamma (\nu + 1) \; \Gamma (\nu + 1/2) } \; {1 \over \sqrt{x}} \;
  \left [\; \exp \left ( + i (x  - {\pi \over 2} (\nu
+{1\over 2})) \right ) \; + \;
        \exp \left ( - i (x -  {\pi \over 2} (\nu + {1\over
2}))\right ) \; \right ] , \nonumber
\end{eqnarray}

\noindent
so when $ j < j ^{2} \ll \epsilon R << \epsilon R $ we derive
\begin{eqnarray}
U^{out}_{run}(R) \sim \left [ { e^{+i\epsilon R} \over
\epsilon R} \left ( C_1 \; \exp (- i {\pi \over 2} (p + {1 \over
2})) \; +\;
        C_2 \; \exp (- i {\pi \over 2}(-p + {1 \over 2}) \right)
\right. \nonumber
\\
\left.  +\; {  e^{-i\epsilon R} \over \epsilon  R} \left ( C_1 \;
\exp (+ i {\pi \over 2} (p + {1 \over 2})) \;+\;
        C_2 \; \exp (+ i {\pi \over 2}(-p + {1 \over 2}))
\right ) \right ]    \; , \label{6.5a}
\end{eqnarray}

\noindent where  $C_1$ and  $C_2$  are given by
\begin{eqnarray}
C_1 =  {\Gamma (a + b + 1 - c) \;\Gamma (1 - c) \over
      \Gamma (b - c + 1) \; \Gamma (a - c + 1)} \;
{2^{-j-1} \; \Gamma (2 p + 1 ) \over (\epsilon R)^{j}\; \Gamma
(p + 1/2)} \; , \nonumber
\\
C_1 = {\Gamma (a + b + 1 - c) \; \Gamma (c - 1) \over \Gamma (a) \;
\Gamma (b) } \; {(\epsilon  R )^{j +1}\; \Gamma (- 2 p  + 1)
\over 2^{-j}\; \Gamma ( p  + 1/2 )} \; . \label{6.5b}
\end{eqnarray}

\noindent The reflection coefficient $R_{\epsilon j}$ is
determined by the coefficients at  $ e^{-i\epsilon R} / \epsilon  R $
and  $ e^{+i\epsilon R} / \epsilon R$.   It is the matter of
simple calculation to verify that when $\epsilon R
\gg j$,  the coefficient $R_{\epsilon j}$ is precisely zero
 \begin{eqnarray}
\epsilon R  \gg j  \;, \qquad R_{\epsilon j}   \equiv 0 \; .
\end{eqnarray}

\noindent
This conclusion  is consistent with the analysis
performed above.

\section{Series expansion on a parameter $R^{-2}$ of  the exact solutions    and
 calculation of the reflection coefficient}

In Section 3 we demonstrate that the basic instructive definition
 for the calculation of  the reflection coefficient in de Sitter model, being grounded exclusively on the use of
 zero order approximation $\Phi^{(0)} (r) $ in the expansion of  a particle wave function in a series
 of the form (\ref{1.1}),  cannot be extended on accounting for
contributions of higher order terms.

Let us start with the solution, the wave running to the horizon
\begin{eqnarray}
U^{out} (z) = \Gamma (a+b-c+1) \; [\; \alpha \; F(z) + \beta \;  G(z) \; ] \; ,
\nonumber
\\
\alpha  = {  \Gamma (1-c) \over \Gamma (b- c +1) \Gamma (a-c +1) } \;,
\qquad
\beta  = {  \Gamma (c-1) \over \Gamma (a) \Gamma (b) } \; ,
\label{A.1}
\end{eqnarray}

\noindent
where
\begin{eqnarray}
F(z) = z^{(p-1/2)/2} (1-z)^{-i\epsilon/2}F(a,b,c;z) \; , \qquad c = j+3/2 = 1+p  \; ,
\nonumber
\\
a = {1 + p  -i \epsilon + i \sqrt{m^{2} -1/4} \over 2}\; ,
\qquad
b = {1+p  -i \epsilon - i \sqrt{m^{2} -1/4} \over 2}\; ,
\label{A.2}
\end{eqnarray}

\noindent
and
\begin{eqnarray}
G(z) = z^{(-p-1/2)/2} (1-z)^{+i\epsilon/2}F(a-c+1,b-c+1 , 2- c;z) \; , \qquad 2- c =  1-p \; ,
\nonumber
\\
a -c+1 = {1 - p  -i \epsilon + i \sqrt{m^{2} -1/4} \over 2}\; ,\qquad
b -c +1 = {1-p  -i \epsilon - i \sqrt{m^{2} -1/4} \over 2}\;.
\label{A.3}
\end{eqnarray}

For the following we need the expressions for all quantities in usual units.
It is convenient to change slightly the designation:
now $R$  stands for the curvature radius
\begin{eqnarray}
z = {r^{2} \over R^{2}} \; ,   \qquad
\epsilon = {E R \over \hbar c} =  \mu {R \over \lambda}
 \;, \qquad E =  \mu \; Mc^{2} \; ,\qquad
 m = {McR \over \hbar} = {R \over \lambda}  \; , \qquad \lambda = {\hbar \over Mc } \; .\qquad \qquad
\label{A.4}
\end{eqnarray}

\noindent
Now, the relations  (\ref{A.1})--(\ref{A.3})  read
\begin{eqnarray}
F(z) =  R^{-p+1/2}  \; {r^{p} \over \sqrt{r} }
\left (1-{ r ^{2}\over R^{2} } \right )^{-i\mu R/ 2 \lambda}
F(a,b,c; \; { r^{2} \over R^{2}}) \; , \qquad c =  1+p  \; , \qquad
\nonumber
\\
a = {1 \over 2}  \left ( 1 + p  -i \mu { R \over  \lambda }  + i
\sqrt{ {R ^{2} \over \lambda^{2} } -{1 \over 4} } \right )  ,
\qquad b = {1 \over 2}  \left ( 1 + p  -i \mu { R \over  \lambda }
- i \sqrt{ {R ^{2} \over \lambda^{2} } -{1 \over 4} } \right )  ;
\label{A.5}
\end{eqnarray}

\noindent and
\begin{eqnarray}
G(z) =  R^{ p+1/2}  \; {r^{-p} \over \sqrt{r} }  \left (1-{ r^{2}  \over R^{2} } \right )^{+ i\mu R/ 2 \lambda }
  F(a-c+1,b-c+1 , 2- c ;\; { r ^{2}\over R^{2}}) \; , \qquad 2- c =  1-p \; ,
\nonumber
\\
a -c+1 = {1 \over 2}  \left ( 1 - p  -i \mu { R \over  \lambda } +
i \sqrt{ {R ^{2} \over \lambda^{2} } -{1 \over 4} } \right )  ,
\;\; b -c +1 = {1 \over 2}  \left ( 1 - p  -i \mu { R \over
\lambda }  - i \sqrt{ {R ^{2} \over \lambda^{2} } -{1 \over 4} }
\right )  . \nonumber
\\
\label{A.6}
\end{eqnarray}

The task consists  in obtaining the approximate expressions for   $F(r)$  and $G(r)$ in the region
far from horizon, $r \ll R$; first  let us preserve the leading  and next to leading terms
(we have a natural small parameter  $\lambda / R$).

First, let us consider the exponential factors
\begin{eqnarray}
\left (1-{ r^{2}  \over R^{2} } \right )^{\pm  i\mu R/ 2 \lambda }
 =   \exp \left [   \pm i{ \mu R \over  2 \lambda }  \ln (1-{ r^{2}  \over R^{2} })  \right ]  =
  \cos \left [  { \mu R \over  2 \lambda }  \ln (1-{ r^{2}  \over R^{2} })  \right ]
 \pm  i \sin \left [  { \mu R \over  2 \lambda }  \ln (1-{ r^{2}  \over R^{2} })  \right ]  .
\nonumber
\\
\end{eqnarray} \label{A.7}

\noindent Using the expansion for the logarithmic function
\begin{eqnarray}
\ln (1 -x) = - ( x + {x^{2} \over 2} + {x^{3} \over 3} + ... ) \; ,\qquad
\ln (1-{ r^{2}  \over R^{2} })= -   \left ({ r^{2}  \over  \; R^{2} }  +
{1 \over 2} { r^{4}  \over  \; R^{4} } + {1 \over 3}  { r^{6}  \over  R^{6} } + ... \right ) \; ,
\nonumber
\end{eqnarray}

\noindent we get  (assuming that  $r^{2} \ll \lambda R$)
\begin{eqnarray}
\left (1-{ r^{2}  \over R^{2} } \right )^{\pm   i\mu R/ 2 \lambda }
=
 \cos  { \mu R \over  2 \lambda }\left ( { r^{2}  \over  \; R^{2} }   +
{1 \over 2} { r^{4}  \over  \; R^{4} } + {1 \over 3}  { r^{6}  \over  R^{6} } + ... \right )
  \nonumber
 \\
 \mp i \sin  { \mu R \over  2 \lambda }   \left ( { r^{2}  \over  \; R^{2} }  +
{1 \over 2} { r^{4}  \over  \; R^{4} } + {1 \over 3}  { r^{6}  \over  R^{6} } + ... \right )
 \approx
  \left ( 1  -   { \mu^{2}  r^{4}  \over  8 \lambda^{2}   R ^{2} } \right )
 \mp   i \;
 { \mu R \over  2 \lambda }   \left ( { r^{2}  \over  \; R^{2} }  +
{1 \over 2} { r^{4}  \over  \; R^{4} } \right ) . \label{A.8}
\end{eqnarray}

\noindent
Terms in eq.~(\ref{A.8}) can be written in  descending order
\begin{eqnarray}
\left (1-{ r^{2}  \over R^{2} } \right )^{\pm   i\mu R/ 2 \lambda } \approx
1 \mp  i \mu   \; {r^{2} \over 2 \lambda R} - {\mu ^{2} \over 2} \;  {r^{4} \over 4\lambda ^{2} R^{2}}
 \mp  i \mu  \; X \;
{r^{4} \over 4 \lambda ^{2} R^{2}} \;,
\nonumber
\\
 X = {\lambda \over R} \ll 1 \;, \qquad { r^{2} \over 2 \lambda R } \ll 1 \; .\qquad \qquad \qquad
\label{A.8'}
\end{eqnarray}

Now we turn to the hypergeometric function
 $F(a,b,c; z)$ from  (\ref{A.5}). Because
 $R \sim 10^{30}, \lambda \sim 10^{-12}$, one can use the
 approximation of a leading and two next order terms in the expressions  for the following parameters
 \begin{eqnarray}
a = {1 \over 2}  \left (
1 + p  -i \mu { R \over  \lambda }  + i { R \over  \lambda }  \sqrt{ 1 -{ \lambda^{2}  \over 4 R^{2} } }
\right ) = { 1 + p \over 2}   -i {\mu -1 \over 2}  { R \over  \lambda }  -i  { \lambda  \over 16 R } \; ,
\nonumber
\\
b = {1 \over 2}  \left (
1 + p  -i \mu { R \over  \lambda }  - i { R \over  \lambda }  \sqrt{ 1 -{ \lambda^{2}  \over 4 R^{2} } }
\right ) = { 1 + p \over 2}   -i {\mu +1 \over 2}  { R \over  \lambda }  + i { \lambda  \over 16 R } \; .
\label{A.9}
\end{eqnarray}

\noindent Then, the hypergeometric function is given as
\begin{eqnarray}
F(a,b,c; \; { r^{2} \over R^{2}})= 1 + {1 \over R^{2}} {ab \over c   }  r^{2} +
{1 \over  2!} {1 \over R^{4}} {a(a+1) b(b+1)  \over c  (c+1)  }  (r^{2})^{2}
\nonumber
\\
+
{1 \over  3!} {1 \over R^{6}} {a(a+1)(a+2) b(b+1)(b+2)  \over c  (c+1) (c+2)   }  (r^{2})^{3} + ...
\nonumber
\\
+ {1 \over  n!} {1 \over R^{2n}} {a(a+1)(a+2)... (a+ n-1)  b(b+1)(b+2).. (b+n-1)  \over c  (c+1) (c+2)...(c+n-1)
   }  (r^{2})^{n} + ...
\label{A.10}
\end{eqnarray}

It is convenient to introduce a shortening notation for small
quantity $X = \lambda / R$), then a typical  term is represented
as
\begin{eqnarray}
{1 \over R^{2}}  {(a+n)(b+n) \over (c +n)  } \approx {1 \over p+1 + n} \times  \qquad \qquad
\nonumber
\\
\times
{-i (\mu -1) \over 2\lambda} \left ( 1 + i {1 + p + 2n \over \mu -1 } \; X +{X^{2} \over 8 (\mu -1)}  \right )
{-i (\mu +1) \over 2\lambda} \left ( 1 + i {1 + p + 2n \over \mu +1 } \; X -{X^{2} \over 8 (\mu +1)}   \right ) \approx
\nonumber
\end{eqnarray}
\begin{eqnarray}
\approx {1 \over p+1 +n} \left (- {\mu^{2} -1 \over 4 \lambda ^{2} } \right ) \left [ 1 +   {2i \mu \over \mu^{2} -1} (1 + p + 2n) \; X -
 { (1+ p+2n)^{2} -1/4 \over \mu^{2} -1}\; X^{2}   \right ] \; .
\end{eqnarray}

\noindent Below we will use the  notation
$$
k^{2} = {\mu^{2} - 1 \over \lambda^{2}} \; ,
$$
then
\begin{eqnarray}
{1 \over R^{2}}  {(a+n)(b+n) \over (c +n)  } \approx {1 \over p+1 +n}
\left (- {k^{2}  \over 4 } \right )
 \left [
 1 +  {2i \mu \over \mu^{2} -1} (1 + p + 2n) \; X -
 { (1+ p+2n)^{2} -1/4 \over \mu^{2} -1}\; X^{2}   \right ]  .
\nonumber
\end{eqnarray}

Thus, we have the following approximate expressions for the first
few terms in the series
 \begin{eqnarray}
{r^{2}  \over R^{2} } {ab \over c }   \approx {1 \over p+1 }
\left (- {k^{2}r^{2}  \over 4 } \right )
\left [  1 +  {2i \mu \over \mu^{2} -1} (1 + p ) \; X -
 { (1+ p)^{2} -1/4 \over \mu^{2} -1}\; X^{2}   \right ]
   ,
\nonumber
\end{eqnarray}
\begin{eqnarray}
{r^{2}  \over R^{2} } {(a+1)(b+1) \over (c+1) }   \approx {1 \over p+ 2 }
\left (- {k^{2} r^{2} \over 4 } \right )
\left [
 1 +  {2i \mu \over \mu^{2} -1} (1 + p +2 \times 1) \; X -
 { (1+ p+2 \times 1 )^{2} -1/4 \over \mu^{2} -1}\; X^{2}   \right ]
   ,
\nonumber
\end{eqnarray}
\begin{eqnarray}
{r^{2}  \over R^{2} } {(a+2)(b+2) \over (c+2) }   \approx {1 \over p+ 3 }
\left (- {k^{2} r^{2} \over 4 } \right )
\left [
 1 +  {2i \mu \over \mu^{2} -1} (1 + p +2 \times 2) \; X -
 { (1+ p+2 \times 2 )^{2} -1/4 \over \mu^{2} -1}\; X^{2}   \right ]
   ,
\nonumber
\end{eqnarray}
\begin{eqnarray}
{r^{2}  \over R^{2} } {(a+3)(b+3) \over (c+3) }   \approx {1 \over p+ 4 }
\left (- {k^{2} r^{2}  \over 4 } \right )
\left [
 1 +  {2i \mu \over \mu^{2} -1} (1 + p +2 \times 3) \; X -
 { (1+ p+2 \times 3 )^{2} -1/4 \over \mu^{2} -1}\; X^{2}   \right ]
   ,
\nonumber
\end{eqnarray}
\begin{eqnarray}
{r^{2}  \over R^{2} } {(a+4)(b+4) \over (c+4) }   \approx {1 \over p+ 5 }
\left (- {k^{2} r^{2}  \over 4 } \right )
\left [
 1 +  {2i \mu \over \mu^{2} -1} (1 + p +2 \times 4) \; X -
 { (1+ p+2 \times 4 )^{2} -1/4 \over \mu^{2} -1}\; X^{2}   \right ]
   ,
\nonumber
\end{eqnarray}

\hspace{5mm}
............................................................................................................

\begin{eqnarray}
{r^{2} \over R^{2}}  {(a+n)(b+n) \over (c +n)  } \approx {1 \over p+1 +n}
\left (- {k^{2} r^{2} \over 4 } \right )
\left [
 1 +  {2i \mu \over \mu^{2} -1} (1 + p + 2n) \; X -
 { (1+ p+2n)^{2} -1/4 \over \mu^{2} -1}\; X^{2}   \right ]  .
\nonumber
\end{eqnarray}

Now we are ready to write down the expressions for the coefficients of the
hypergeometric series preserving only leading and two next order
terms
 \begin{eqnarray}
{1 \over 1!} {r^{2}  \over R^{2} } {ab \over c }   \approx {1 \over p+1 }
\left (- {k^{2}r^{2}  \over 4 } \right ) \left [
 1 +  {2i \mu \over \mu^{2} -1} (1 + p ) \; X -
 { (1+ p)^{2} -1/4 \over \mu^{2} -1}\; X^{2}   \right ]
   ,
\nonumber
\end{eqnarray}

\begin{eqnarray}
{1 \over 2!} {(r^{2})^{2} \over (R^{2})^{2} } {ab  (a+1) (b+1) \over c (c+1)  }  \approx \hspace{30mm}
\nonumber
\\
\approx
\left (- {k ^{2} r^{2}  \over 4 }\right  )^{2}  {1  \over 2!(p+1)(p+2)}
\left \{   1 +  {2i \mu \over \mu^{2} -1} [ (1 + p )  + (1+ p + 2 \times 1) ] \; X - \right.
\nonumber
\\
\left. - X^{2}
\left [  {4\mu^{2} \over (\mu^{2} -1)^{2} } (1+p) (1 + p + 2 \times 1) +
{(1+p)^{2} -1/4 \over \mu^{2} - 1} +  {(1+p + 2 \times 1)^{2} -1/4 \over \mu^{2} - 1} \right ]
 \right \} \; ,
\nonumber
\end{eqnarray}

\begin{eqnarray}
{1 \over 3!} {( r^{2})^{3} \over (R^{2})^{3} } {ab  (a+1) (b+1) (a+2) (b+2) \over c (c+1) (c+2)  }  \approx
\left (- {k ^{2} r^{2}  \over 4 }\right  )^{3}   {1  \over 3!(p+1)(p+2)(p+3)} \times
\nonumber
\\
\times
\left \{   1 + X\;
 {2i \mu \over \mu^{2} -1} [ (1 + p )  + (1+ p + 2 \times 1)+    (1+ p + 2 \times 2)] - \right.
  \nonumber
 \\
   - X^{2} \left [ {4\mu^{2} \over (\mu^{2}-1)^{2}} [ (1+p) + (1+p + 2 \times 1)] (1 + p +2 \times 2 )
 + \right.
 \nonumber
 \\
 \left.   \left.  +
 {(1+p)^{2} -1/4 \over \mu^{2} - 1} +  {(1+p + 2 \times 1)^{2} -1/4 \over \mu^{2} - 1} +
{(1+ p +2 \times 2)^{2} -1/4 \over \mu^{2} - 1}   \right ]
 \right \} \; ,
\nonumber
\end{eqnarray}

\begin{eqnarray}
{1 \over 4!} {(r^{2})^{4} \over (R^{2} )^{4} } {ab  (a+1) (b+1) (a+2) (b+2) (a+3) (b+3)\over c (c+1) (c+2)(c+3)   }
\approx \qquad \qquad \qquad \qquad
\nonumber
\\
\approx
\left (- {k ^{2} r^{2}  \over 4 }\right  )^{4}  {1   \over 4!(p+1)(p+2)(p+3)(p+4) } \times
\qquad \qquad \qquad \qquad \qquad
\nonumber
\\
\times
\left \{   1 + X \;  {2i \mu \over \mu^{2} -1} [ (1 + p )  + (1+ p + 2 \times 1)
+ (1+ p + 2 \times 2) +  (1+ p + 2 \times 3 ) ]  -
\right. \qquad \qquad
  \nonumber
 \\
   - X^{2} \left [ {4\mu^{2} \over (\mu^{2}-1)^{2}} [ (1+p) + (1+p + 2 \times 1) + (1+ p + 2 \times 2) ]
    (1 + p +2 \times 3 )
 + \right. \qquad \qquad
 \nonumber
 \\
 \left.   \left.  +
 {(1+p)^{2} -1/4 \over \mu^{2} - 1} +  {(1+p + 2 \times 1)^{2} -1/4 \over \mu^{2} - 1} +
{(1+ p +2 \times 2)^{2} -1/4 \over \mu^{2} - 1}   +  {(1+ p +2 \times 3)^{2} -1/4 \over \mu^{2} - 1} \right ]
 \right \} \; ,
\nonumber
\end{eqnarray}

.......................................................................................................................................

\begin{eqnarray}
{1 \over n!} {(r^{2})^{n}  \over (R^{2})^{n} }
 { {ab \;  (a+1) (b+1) .... (a+n-1) (b+n-1) \over c (c+1) ... (c+n) }    }  \approx
\left (- {k^{2} r^{2}  \over 4 }  \right )^{n} {  1  \over n!(p+1)(p+2) ... (p+n) } \times \qquad \qquad \qquad\qquad
\nonumber
\end{eqnarray}
\begin{eqnarray}
\times
\left \{   1 + X\; {2i \mu \over \mu^{2} -1} \left  [ (1 + p )  + (1+ p + 2 \times 1)
+  (1+ p + 2 \times 2) + ... + (1+ p + 2 \times (n-1))  \right ] -
\right.
  \nonumber
 \\
   - X^{2} \left [ {4\mu^{2} \over (\mu^{2}-1)^{2}} [ (1+p) + (1+p + 2 \times 1) +  ... +
   (1+ p + 2 \times (n-2) )  ]\;
    (1 + p +2 \times (n-1) )
 + \right.
 \nonumber
 \\
 \left.   \left.  +
 {(1+p)^{2} -1/4 \over \mu^{2} - 1} +  {(1+p + 2 \times 1)^{2} -1/4 \over \mu^{2} - 1} +
...+
 {(1+ p +2 \times (n-1))^{2} -1/4 \over \mu^{2} - 1} \right ]
 \right \} \; . \qquad \qquad
\nonumber
\end{eqnarray}

Thus, initial exact hypergeometric function can be approximated
by the sum of three series
\begin{eqnarray}
\bar{F} = F(a,b,c; \; { r^{2} \over R^{2}}) = \bar{F}_{0} (r) + X  \; \bar{F}_{1}(r)
+  X^{2}  \; \bar{F}_{2}(r)
\; .
\label{A.13}
\end{eqnarray}

The leading series   $\bar{F}_{0} (r) $, in fact, reduces to the Bessel function
\begin{eqnarray}
\bar{F}_{0} (r)=    1 +  {  (i  kr /2)^{2}   \over n! (p+1) }
+ {  (i  kr /2)^{4}   \over 2! (p+1)(p+2) } + ... +   {  (i  kr /2)^{2n}   \over n! (p+1)(p+2) ... (p+n) }  + ...
      \nonumber
   \\
   =  \Gamma (p+1) \sum_{n=0}^{\infty}
   {  (i  kr /2)^{2n}   \over n! \Gamma(p+1 +n ) } =
   \Gamma (1+p) \left ( {kr \over 2}    \right )^{-p} J_{p} (kr) \; ,
   \qquad \qquad
   \label{A.14}
\end{eqnarray}

\noindent
where
$$
J_{p}(x) = ( {x \over 2} )^{p} \sum_{0}^{\infty}  { (-x^{2} /4)^{n}
\over  n! \; \Gamma (p+1 +n) } \; , \qquad  x = kr \; .
$$

The second series is given by
\begin{eqnarray}
X\; \bar{F}_{1}(r) =  X\;  {2i \mu \over \mu^{2} -1}  \left \{
 (- k^{2} r^{2} / 4  ) {p+1  \over p +1}    +
  (- k^{2} r^{2} / 4  )^{2}   {[ (1 + p )  + (1+ p + 2 \times 1) ] \over2! (p +1)(p+2)} + \right.
  \nonumber
  \\
  +
   (- k^{2} r^{2} / 4  )^{3}   {[ (1 + p )  + (1+ p + 2 \times 1)  +   (1+ p + 2 \times 2)] \over3!(p +1)(p+2)(p+3) }
+
\nonumber
\\
+ (- k^{2} r^{2} / 4  )^{4}   {[ (1 + p )  + (1+ p + 2 \times 1)  +   (1+ p + 2 \times 2)
+   (1+ p + 2 \times 3) ] \over 4! (p +1)(p+2)(p+3)(p+4)  } + ...
\nonumber
\\
\left.
(- k^{2} r^{2} / 4  )^{n}   {[ (1 + p )  + (1+ p + 2 \times 1)  + ... +
(1+ p + 2 \times (n-1)) ] \over n! (p +1)(p+2)... (p+n)  }
\right \} \; ;
\label{A.15}
\end{eqnarray}

\noindent
that is
\begin{eqnarray}
X \bar{F}_{1}(r) =  X  {2i \mu \over \mu^{2} -1} \left  ({ - k^{2} r^{2} \over  4 } \right )  \left \{
  1     +
  \left  ({ - k^{2} r^{2} \over  4 } \right )
    {[ (1 + p )  + (1+ p + 2 \times 1) ] \over2! (p +1)(p+2)} + \right.
  \nonumber
  \\
  +
   \left  ({ - k^{2} r^{2} \over  4 } \right ) ^{2}   {[ (1 + p )  + (1+ p + 2 \times 1)  +   (1+ p + 2 \times 2)] \over3!(p +1)(p+2)(p+3) }
+
\nonumber
\\
+ \left  ({ - k^{2} r^{2} \over  4 } \right )   ^{3}   {[ (1 + p )  + (1+ p + 2 \times 1)  +   (1+ p + 2 \times 2)
+   (1+ p + 2 \times 3) ] \over 4! (p +1)(p+2)(p+3)(p+4)  } + ...
\nonumber
\\
\left.  ... +
 \left  ({ - k^{2} r^{2} \over  4 } \right )   ^{n}   {[ (1 + p )  + (1+ p + 2 \times 1)  + ... +
(1+ p + 2 \times n ) ] \over (n+1)! (p +1)(p+2)... (p+n +1 )  }
\right \} \; ;
\end{eqnarray}

\noindent
or shorter
\begin{eqnarray}
X \bar{F}_{1}(r)
= X {2i \mu \over \mu^{2} -1}  \;  \Gamma(p+1)   \left ( {- k^{2} r^{2} \over  4 }   \right ) \;\;
\sum_{n=0}^{\infty}
 \left  ({ - k^{2} r^{2} \over  4 } \right )   ^{n}  \;  {[ (1 + p )    + ...+
    (1+ p + 2 \times n) ] \over (n+1) ! \;  \Gamma (p+ 2 + n)  }\; .
\nonumber
\\
\label{A.16}
\end{eqnarray}

Using known sums
$$
[ (1 + p )    + ...+    (1+ p + 2 \times n) = (1+p)(n+1) + 2 (1 + 2 +3 + ...  n) =
$$
$$
 = (p+1) (1+ n) +2 {(1+n) n \over 2} = (n+1) (n+1+p)
$$

\noindent
we derive
\begin{eqnarray}
X \bar{F}_{1}(r)
= X {2i \mu \over \mu^{2} -1}  \;  \Gamma(p+1)   \left ( {- k^{2} r^{2} \over  4 }   \right ) \;\;
\sum_{n=0}^{\infty}
 (- k^{2} r^{2} / 4  )^{n}  \;  { (n+1) (n+1+p) \over (n+1) ! \;  \Gamma (p+ 2 + n)  }=
\nonumber\\
= X {2i \mu \over \mu^{2} -1}  \;  \Gamma(p+1)   \left ( {- k^{2} r^{2} \over  4 }   \right ) \;\;
\sum_{n=0}^{\infty}
   \;  {  (- k^{2} r^{2} / 4  )^{n} \over n ! \;  \Gamma (p+ 1 + n)  }= \qquad \qquad
   \nonumber
   \\
   =  X \left ( {- k^{2} r^{2} \over  4 }   \right )   {2i \mu \over \mu^{2} -1}  \;
    \Gamma(p+1)  \left ({kr \over 2} \right )^{-p}   \;\;
J_{p} (kr) = X
\left ( {- k^{2} r^{2} \over  4 }   \right )   {2i \mu \over \mu^{2} -1} \;  \bar{F}_{0}(r)
 \end{eqnarray}

Thus, the approximation (\ref{A.13})  can be presented as follows
\begin{eqnarray}
\bar{F} =  \bar{F}_{0} (r) + X  \bar{F}_{1}(r) + X^{2}  \bar{F}_{2}(r) =
\bar{F}_{0} (r)  +  X  \;
   {2i \mu \over \mu^{2} -1}  \;  ( {- k^{2} r^{2} \over  4 }     ) \; \bar{F}_{0} (r) +  X^{2}  \bar{F}_{2}(r)
\; ,
\label{A.13'}
\end{eqnarray}

Similar relations can be  derived  for the
hypergeometric series   $\bar{G}(r)$:
\begin{eqnarray}
\bar{G} = F(a-c+1 ,b-c+1, 2- c; \; { r^{2} \over R^{2}}) =
\bar{G}_{0} (r) + X \; \bar{G}_{1}(r)
+ X^{2}  \; \bar{G}_{2}(r) \; .
\label{A.17}
\end{eqnarray}

\noindent
The leading term again reduces to the Bessel function
\begin{eqnarray}
\bar{G}_{0} (r)=    \Gamma (1-p) \left ( {kr \over 2}
   \right )^{+p} J_{-p} (kr) \; . \qquad \qquad
   \label{A.18}
\end{eqnarray}

\noindent
Next order term is given by
\begin{eqnarray}
X \bar{G}_{1}(r)=  X
\left ( {- k^{2} r^{2} \over  4 }   \right )   {2i \mu \over \mu^{2} -1}   \bar{G}_{0}(r)
 \end{eqnarray}

So, the approximation  (\ref{A.17}) is presented as
\begin{eqnarray}
\bar{G} =  \bar{G}_{0} (r) + X  \bar{G}_{1}(r) + X^{2}  \bar{G}_{2}(r) =
\bar{G}_{0} (r)  +  X  \;
   {2i \mu \over \mu^{2} -1}  \;  ( {- k^{2} r^{2} \over  4 }     ) \; \bar{G}_{0} (r)
    +  X^{2}  \bar{G}_{2}(r)
\; ,
\label{A.17'}
\end{eqnarray}

Now let us consider
the whole function
\begin{eqnarray}
F(z) =  R^{-p+1/2}  \; {r^{p} \over \sqrt{r} }
\left (1-{ r ^{2}\over R^{2} } \right )^{-i\mu R/ 2 \lambda}
F(a,b,c; \; { r^{2} \over R^{2}})  =
\nonumber
\\
=
R^{-p+1/2} \; {r^{p} \over \sqrt{r} }
\left [  1 +  i \mu   \; {r^{2} \over 2 \lambda R} - {\mu ^{2} \over 2} \;  {r^{4} \over 4\lambda ^{2} R^{2}}
 +  i \mu  \; X \; {r^{4} \over 4 \lambda ^{2} R^{2}} \right ] \times
\nonumber
\\
\times
\left [
\bar{F}_{0} (r)  +
   {2i \mu \over \mu^{2} -1}  \;  ( {- k^{2} r^{2} \over  4 } )   \; X  \bar{F}_{0} (r)  +  X^{2}  \bar{F}_{2}(r)
\right ] =
\nonumber
\end{eqnarray}

\begin{eqnarray}
=
R^{-p+1/2}  \; {r^{p} \over \sqrt{r} }
\left [
\bar{F}_{0} (r)  +
   {2i \mu \over \mu^{2} -1}  \;  ( {- k^{2} r^{2} \over  4 } )   \; X  \bar{F}_{0} (r)  +  X^{2}  \bar{F}_{2}(r) +
\right.
\nonumber
\end{eqnarray}

\begin{eqnarray}
+
  i \mu   \; {r^{2} \over 2 \lambda R}  \bar{F}_{0} (r)     +
  i \mu   \; {r^{2} \over 2 \lambda R}
   {2i \mu \over \mu^{2} -1}  \;  ( {- k^{2} r^{2} \over  4 } )   \; X  \bar{F}_{0} (r)    +
   i \mu   \; {r^{2} \over 2 \lambda R}  X^{2}  \bar{F}_{2}(r)
  -
\nonumber
\end{eqnarray}
\begin{eqnarray}
 - {\mu ^{2} \over 2} \;  {r^{4} \over 4\lambda ^{2} R^{2}} \bar{F}_{0} (r)
 - {\mu ^{2} \over 2} \;  {r^{4} \over 4\lambda ^{2} R^{2}}
   {2i \mu \over \mu^{2} -1}  \;  ( {- k^{2} r^{2} \over  4 } )
    \; X  \bar{F}_{0} (r)  - {\mu ^{2} \over 2} \;  {r^{4} \over 4\lambda ^{2} R^{2}}  X^{2}  \bar{F}_{2}(r)
  +
\nonumber
\end{eqnarray}
\begin{eqnarray}
\left. +
i \mu  \; X \; {r^{4} \over 4 \lambda ^{2} R^{2}} \bar{F}_{0} (r)  +
 i \mu  \; X \; {r^{4} \over 4 \lambda ^{2} R^{2}}
   {2i \mu \over \mu^{2} -1}  \;  ( {- k^{2} r^{2} \over  4 } )   \; X  \bar{F}_{0} (r)  +
   i \mu  \; X \; {r^{4} \over 4 \lambda ^{2} R^{2}}  X^{2}  \bar{F}_{2}(r)
\right ]
\nonumber
\end{eqnarray}

\begin{eqnarray}
=
R^{-p+1/2}  \; {r^{p} \over \sqrt{r} }
\left [
\bar{F}_{0} (r)  -  i \mu \; { r^{2} \over 2 \lambda R} \;  \bar{F}_{0} (r)   +  X^{2}  \bar{F}_{2}(r) +
\right.
\nonumber
\end{eqnarray}

\begin{eqnarray}
+  i \mu  {r^{2} \over 2 \lambda R}  \bar{F}_{0} (r)
+ \mu^{2} ( {r^{2} \over 2 \lambda R})^{2}   \bar{F}_{0} (r)   +
   i \mu   \; {r^{2} \over 2 \lambda R}  \; X^{2}  \bar{F}_{2}(r)
  -
\nonumber
\end{eqnarray}
\begin{eqnarray}
 -  {\mu ^{2} \over 2} \; ( {r^{2} \over 2\lambda  R})^{2} \;  \bar{F}_{0} (r)
 + {i\mu^{3} \over 2} ({r^{2} \over 2\lambda R})^{3}  \;
   \bar{F}_{0} (r)  - {\mu ^{2} \over 2} \;  ( {r^{2} \over 2\lambda  R})^{2} \;   X^{2}  \bar{F}_{2}(r)
  +
\nonumber
\end{eqnarray}
\begin{eqnarray}
\left. +
i \mu  \; X \; ({r^{2} \over 2 \lambda  R})^{2} \;  \bar{F}_{0} (r)  +
\mu^{2} ({r^{2} \over 2 \lambda R})^{3}  \bar{F}_{0} (r)  +
   i \mu  \;  ({r^{2} \over 2 \lambda  R})^{2}   X^{3}  \bar{F}_{2}(r)
\right ]
\nonumber
\end{eqnarray}

Preserving only first two   terms we have
\begin{eqnarray}
F(z) =
R^{-p+1/2}  \; {r^{+p} \over \sqrt{r} }
\left [
\bar{F}_{0} (r)     + {1 \over 8} \mu^{2}  {r^{2} r^{2} \over  \lambda^{2}  R^{2}} \;   \bar{F}_{0} (r)
+   {\lambda ^{2} \over R^{2}}   \bar{F}_{2}(r)
  \right ] .
  \label{A}
\end{eqnarray}

\noindent
Similarly, for  $G(r)$ we obtain
\begin{eqnarray}
G (z) =
R^{+p+1/2}  \; {r^{-p} \over \sqrt{r} }
\left [
\bar{G}_{0} (r)     + {1 \over 8} \mu^{2}  {r^{2} r^{2}  \over  \lambda^{2}  R^{2}} \;   \bar{G}_{0} (r)
+   {\lambda ^{2} \over R^{2}}   \bar{G}_{2}(r)
  \right ] .
 \label{A'}
 \end{eqnarray}

\noindent
One should emphasize one feature of  the expansions
(\ref{A}) and  (\ref{A'}):  these approximations are real valued as we must expect remembering on relations
(\ref{3.3c}) and (\ref{3.5d}).

In the known method of determining and calculating the
reflection coefficients in de Sitter model [3--6], authors  used the only  leading terms
in approximations (\ref{A}) and  (\ref{A'}), because only   these terms allows
to separate elementary solutions of the form
 $  e^{ \pm i kr }  $ -- see (\ref{6.5a}).

Let us perform some  additional calculations to clarify the problem.
The functions $F(z)$ and $F(z)$ enter the expression for  (to horizon) running  wave
\begin{eqnarray}
U^{out} (z) = \Gamma (a+b-c+1) [ \; \alpha \; F(z) + \beta \;  G(z) \; ] \; ,
\nonumber
\\
\alpha  = {  \Gamma (1-c) \over \Gamma (b- c +1) \Gamma (a-c +1) } \;,
\qquad
\beta  = {  \Gamma (c-1) \over \Gamma (a) \Gamma (b) } \; ,
\label{A.25}
\end{eqnarray}

\noindent
where  (introducing a very large parameter   $Y = R /  2\lambda $)
\begin{eqnarray}
\alpha \approx  { \Gamma (-p) \over
\Gamma [ -i (\mu -1)  Y  + (1-p) /2  )] \;\;
\Gamma [ -i (\mu +1)  Y  + (1-p) /2  ) ] } \; ,
\nonumber
\end{eqnarray}
\begin{eqnarray}
\beta  \approx  { \Gamma (+p) \over
\Gamma [ -i (\mu -1) Y  + (1+p) /2  )] \;\;
\Gamma [ -i (\mu +1) Y  + (1+p) /2  ) ] } \; .
\label{A.26}
\end{eqnarray}

For all physically reasonable  values of quantum numbers $j$ (not very high ones)  and values of
 $\mu$ (different  from the critical value $\mu  = 1$ and not too high ones -- they correspond  in fact to energies
 of a particle in
 units of the rest energy) the argument of $\Gamma$-functions in (\ref{A.26}) are complex-valued with
 very large imaginary parts.

Let us  multiply the given solution  (\ref{A.25}) by  a special factor $A$ which permits us to distinguish
small and large parts in this expansion:
\begin{eqnarray}
A = {\Gamma (a -p/2 + 1/4 )  \Gamma (b -p/2 +1/4 )  \over
\Gamma (a + b -c +1  ) }\;.
\label{A.27}
\end{eqnarray}

\noindent
So, instead of  (\ref{A.25}), we  get
\begin{eqnarray}
A U^{out} (z) =  \; \alpha  ' \; F(z) + \beta ' \;  G(z) \; ,
\nonumber
\\
\alpha '  =  \Gamma (1-c) {  \Gamma (a -p/2 +1/4 )  \Gamma (b -p/2 +1/4)  \over \Gamma (b- c +1) \Gamma (a-c +1) } \;,
\nonumber
\\
\beta '   =  \Gamma (c-1)  {    \Gamma (a -p/2+1/4 )  \Gamma (b -p/ +1/4) \over \Gamma (a) \Gamma (b) } \;.
\label{A.28}
\end{eqnarray}

\noindent
Instead of (\ref{A.26}),  the expressions for $\alpha  '$    and $ \beta ' $  are
\begin{eqnarray}
\alpha ' \approx   \Gamma (-p) \;
{ \Gamma [ -i (\mu -1) Y  + 1/4  ]  \over  \Gamma [ -i (\mu -1)  Y  + (1-p) /2  )] } \;\;
 { \Gamma [ -i (\mu +1) Y  + 1/4   ] \over \Gamma [ -i (\mu +1)  Y  + (1-p) /2  ) ] }\; ,
\nonumber
\\
\beta '  \approx  \Gamma (+p)\;
{ \Gamma [ -i (\mu -1) Y  + 1/4  ] \over  \Gamma [ -i (\mu -1) Y  + (1+p) /2  )] }
\;\; {\Gamma [ -i (\mu +1) Y  + 1/4   ] \over
 \Gamma [ -i (\mu +1) Y  + (1+p) /2  ) ] }\;.
\label{A.29}
\end{eqnarray}

\noindent
Allowing for identities
\begin{eqnarray}
\Gamma(-p)   =  - {\pi \over \sin p \pi } {1 \over  \Gamma (1+p) } \; ,
\qquad
\Gamma(p)   =  + {\pi \over \sin p \pi } {1 \over  \Gamma (1-p) }\; ,
\label{A.30}
\end{eqnarray}

\noindent $\alpha  '$    and $ \beta ' $  (\ref{A.29}) are   transformed into
\begin{eqnarray}
\alpha ' \approx   - {\pi \over \sin p \pi } {1 \over  \Gamma (1+p) }  \;
{ \Gamma [ -i (\mu -1) Y  + 1/4  ]  \over  \Gamma [ -i (\mu -1)  Y  + (1-p) /2  )] } \;\;
 { \Gamma [ -i (\mu +1) Y  + 1/4   ] \over \Gamma [ -i (\mu +1)  Y  + (1-p) /2  ) ] }\; ,
\nonumber
\\
\beta '  \approx  + {\pi \over \sin p \pi } {1 \over  \Gamma (1-p) }\;
{ \Gamma [ -i (\mu -1) Y  + 1/4  ] \over  \Gamma [ -i (\mu -1) Y  + (1+p) /2  )] }
\;\; {\Gamma [ -i (\mu +1) Y  + 1/4   ] \over
 \Gamma [ -i (\mu +1) Y  + (1+p) /2  ) ] } \; .
\label{A.31}
\end{eqnarray}

Now we have to take into account the asymptotic formula for $\Gamma$-function
%  (B-E. Tom 1, P. 62)
\begin{eqnarray}
{\Gamma (z + A ) \over \Gamma (z + B) } =
z^{A-B} \left ( 1 + {1 \over z}  \; {(A-B)(A+B+1) \over 2} + ... \right )\;,\qquad
\mid  \mbox{arg} \; z \mid < \pi \;, \qquad \mid z \mid \rightarrow \infty \; ;
\label{A.32}
\end{eqnarray}

\noindent
then  (remembering that  $ Y = R / 2 \lambda$)
\begin{eqnarray}
{ \Gamma [ -i (\mu -1) Y  + 1/4  ]  \over  \Gamma [ -i (\mu -1)  Y  + (1-p) /2  )] } \approx
\left [ -i {(\mu -1)  \over 2 \lambda } R  \right ]^{p/2-1/4 } \;
\left [ 1 + {2 \lambda  \over  -i (\mu -1) R } { (2p -1)( 7 -2p)   \over 32 } \right  ] \; ,
\nonumber
\end{eqnarray}
\begin{eqnarray}
{ \Gamma [ -i (\mu +1) Y  + 1/4  ]  \over  \Gamma [ -i (\mu +1)  Y  + (1-p) /2  )] } \approx
\left [ -i {(\mu +1)  \over 2 \lambda } R  \right ]^{p/2-1/4}
\left [ 1 + {2 \lambda  \over  -i (\mu +1) R } { (2p -1)( 7 -2p)   \over 32 }   \right  ]  ,
\label{A.33}
\end{eqnarray}

\noindent
and
\begin{eqnarray}
{ \Gamma [ -i (\mu -1) Y  + 1/4  ]  \over  \Gamma [ -i (\mu -1)  Y  + (1+p) /2  )] } \approx
\left [ -i {(\mu -1)  \over 2 \lambda } R  \right ]^{-p/2 -1/4} \; \left [ 1 + {2 \lambda  \over  -i
(\mu -1) R } { (- 2p -1)( 7 +2p)   \over 32 }   \right  ] \; ,
\nonumber
\end{eqnarray}
\begin{eqnarray}
{ \Gamma [ -i (\mu +1) Y  + 1/4  ]  \over  \Gamma [ -i (\mu +1)  Y  + (1+p) /2  )] } \approx
\left [ -i {(\mu +1)  \over 2 \lambda } R  \right ]^{-p/2-1/4} \; \left [ 1 + {2 \lambda  \over  -i (\mu +1) R }
 { (- 2p -1)( 7 +2p)   \over 32 }
   \right  ] \; .
\label{A.34}
\end{eqnarray}

Substituting  (\ref{A.33}) and  (\ref{A.34})  into (\ref{A.31}),  we obtain
\begin{eqnarray}
\alpha ' \approx   -  \left ( { \mu^{2} - 1 \over 4 \lambda ^{2}} R^{2} \right ) ^{p/2-1/4} \; (-1)^{p/2 -1/4} \;
{\pi \over \sin p \pi } {1 \over  \Gamma (1+p) }  \;\times
\nonumber
\\
\times
\left [ 1 + {2 \lambda  \over  -i (\mu -1) R } { (2p -1)( 7 -2p)   \over 32 }  \right  ]
\left [ 1 + {2 \lambda  \over  -i (\mu +1) R } { (2p -1)( 7 -2p)   \over 32 }   \right  ],
\nonumber
\end{eqnarray}
\begin{eqnarray}
\beta '  \approx  + \left ( { \mu^{2} - 1 \over 4 \lambda ^{2}} R^{2} \right ) ^{-p/2-1/4}
\; (-1)^{- p/2 -1/4} \;
{\pi \over \sin p \pi } {1 \over  \Gamma (1-p) }\;\;\times
\nonumber
\\
\times
\left [ 1 + {2 \lambda  \over  -i (\mu -1) R } { (-2p -1)( 7 +2p)   \over 32 }   \right  ]
\left [ 1 + {2 \lambda  \over  -i (\mu +1) R } { (-2p -1)( 7 +2p)   \over 32 }  \right  ].
\label{A.35}
\end{eqnarray}

\noindent
Preserving the only terms  of first two orders
we  have
\begin{eqnarray}
\alpha ' \approx   - {\pi \over \sin p \pi }  {1 \over  \Gamma (1+p) }
 \left ( { k \over 2} \right ) ^{+p-1/2} R^{+p-1/2}    (-1)^{p/2 -1/4}
\left ( 1 + i { (2p -1)( 7 -2p)   \over 8 }   {\mu \over \mu^{2} -1 }  {\lambda \over R} \right ),
\nonumber
\end{eqnarray}
\begin{eqnarray}
\beta ' \approx
+ {\pi \over \sin p \pi } {1 \over  \Gamma (1-p) }
 \left ( { k \over 2} \right ) ^{-p-1/2} R^{-p-1/2}  (-1)^{-p/2 -1/4}
 \left ( 1 - i { (-2p -1)( 7 +2p)   \over 8 }  {\mu \over \mu^{2} -1 }  {\lambda \over R} \right ).
\nonumber
\\
\label{A.36}
\end{eqnarray}

Substituting expressions (\ref{A.36}) into the following
expansion
\begin{eqnarray}
A U^{out} (z) =  \alpha  ' \; F(z) + \beta ' \;  G(z)  \; ,\qquad
\nonumber
\\
F(r) =  R^ {-p+1/2}\;
\Gamma (1+p)   \left (   { k \over 2}    \right ) ^{-p}
 { 1 \over  \sqrt{r} }   J_{p} (kr) \; ,
 \nonumber
\\
G(r) \approx
R^{p+1/2}
\; \Gamma (1-p) \left ( {k \over 2}    \right )^{p}  {1 \over \sqrt{r} } J_{-p} (kr)
\; ,
\label{A.38}
\end{eqnarray}

\noindent we arrive at
the following zero-order  approximation
$$
A U^{out} (z) = \psi^{out}_{0} (r)=
$$
$$
= - {\pi \over \sin p \pi }  {1 \over  \Gamma (1+p) }
 \left ( { k \over 2} \right ) ^{+p-1/2 } R^{+p-1/2 }  (-1)^{p/2 -1/4}
 R^ {-p+1/2}
\Gamma (1+p)   \left (   { k \over 2}    \right ) ^{-p}
 { 1 \over  \sqrt{r} }   J_{p} (kr) +
$$
$$
+
{ \pi \over \sin p \pi } {1 \over  \Gamma (1-p) }
 \left ( { k \over 2} \right ) ^{-p-1/2} R^{-p-1/2} (-1)^{-p/2 -1/4}
R^{p+1/2}
 \Gamma (1-p) \left ( {k \over 2}    \right )^{p}  {1 \over \sqrt{r} } J_{-p} (kr)\; ,
 $$

\noindent
that is
$$
A U^{out} (z) = \psi^{out}_{0} (r)=
$$
\begin{eqnarray}
=  {\pi   \over \sin p \pi } \;\;
    \sqrt{2k \over r }  \left [\;  \; -(-1)^{p/2 -1/4} \;    J_{p} (kr) + \; (-1)^{-p/2 -1/4} \;
 J_{-p} (kr) \;  \right ] =
 \nonumber
 \\
 = - (-1)^{-p/2 -1/4} \; \sqrt{2k \over r } \; \;\;  {\pi   \over \sin p \pi }
 \left [\;  \; (-1)^{p} \;    J_{p} (kr) -
 J_{-p} (kr) \;  \right ] \; ;
\label{A.39}
\end{eqnarray}

\noindent
which coincides with  a spherical wave propagating  in Minkowski space from the origin,
expressed  through the Hankel functions of the first kind
\begin{eqnarray}
H^{(1)}_{p} ={ ip \over \sin p \pi }  \left [  (-1)^{p}  J_{p} (kr)  -  J_{-p}(kr) \right ] \;.
\label{A.40}
\end{eqnarray}

\section{Conclusion}

The last but not the least mathematical remark  should be given.
 All known quantum
mechanical problems with potentials containing one barrier reduce
to a  second order differential equation with four singular
points, the equation of Heun class. In particular, the most
popular cosmological problem of that type is a particle in the
Schwarzschild space-time background and it reduces to the Heun
differential equation. Quantum mechanical problems of  tunneling
type are  never linked  to differential equation of hypergeometric
type, equation with three singular points; but in the case of de Sitter model
the wave equations for different fields, of spin 0, 1/2, and 1, after separation of  variables are reduced to the
second order differential equation with three singular points,  and there  exists no ground to
search  in these  systems  problems of  tunneling class.

\section*{Acknowledgements}

Authors are grateful to participants of seminar of Laboratory of
theoretical physics, Institute of Physics of National Academy of
Sciences of Belarus for stimulating discussion.

 We wish to thank the Organizers of the
 XLVIII All-Russia conference on problems in
Particle Physics, Plasma Physics, Condensed Matter, and Optoelectronics
Russia, Moscow, 15-18 May 2012, dedicated to the 100-th anniversary of Professor Ya.P. Terletsky,
 for  opportunity to give a talk on the subject.

\end{document}